\documentclass[prl, twocolumn,showpacs,preprintnumbers,amsmath,amssymb]{revtex4}
\usepackage{dcolumn}
\usepackage{bm}
\usepackage{natbib}
\usepackage{color} 
\usepackage[normalem]{ulem}
\usepackage{cancel}
\usepackage{graphicx}

\begin{document}

\title{Highly-efficient source of collimated multi-MeV photons driven by radiation\\reaction of an electron beam in a self-generated magnetic field}

\author{D. J. Stark$^1$}
\author{T. Toncian$^2$}
\author{A. V. Arefiev$^1$}

\affiliation{$^1$ Institute for Fusion Studies, The University of Texas, Austin, Texas 78712, USA}
\affiliation{$^2$ Center for High Energy Density Science, The University of Texas at Austin, Texas 78712, USA}

\date{\today}

\begin{abstract}
The rapid development of high brilliance X-ray radiation sources is revolutionizing physics, chemistry, and biology research through their novel applications. Another breakthrough is anticipated with the construction of next-generation laser facilities which will operate at intensities beyond $10^{23}$ $\mathrm{W/cm^2}$,  leading to higher yield, shorter wavelength radiation sources. We use numerical simulations to demonstrate that a source of collimated multi-MeV photons with conversion efficiency comparable to the one expected for these facilities is achievable at an order of magnitude lower in intensity, within reach of the existing facilities. In the optimal setup, the laser pulse irradiates a bulk solid-density target, heating the target electrons and inducing relativistic transparency. As the pulse then propagates, it generates a beam of energetic electrons which in turn drives a strong azimuthal magnetic field. This field significantly enhances the radiation reaction for the electrons, yielding tens of TW of directed MeV photons for a PW-class laser.
\end{abstract}

\maketitle

The rapid improvement in ultra-intense laser pulses has unlocked new areas of physics, both in fundamental research and technological applications. The prospect of generating copious quantities of multi-MeV photons in laser-target interactions has recently attracted particular interest due to its many potential applications, including pair production \cite{Hui}, laboratory astrophysics \cite{labast}, photo-nuclear spectroscopy~\cite{Schreiber,Kwan}, radiation therapy~\cite{Weeks}, and radiosurgery~\cite{Girolami}.

Presently, Compton backscattering is one of the primary means for gamma-ray generation, combining conventional electron acceleration with laser technology~\cite{Chen}. Several laser facilities are due to be commissioned in the next few years that are expected to operate at intensities beyond $10^{23}$ $\mathrm{W/cm^2}$ and that will potentially enable all-optical gamma-ray sources \cite{Phuoc,Powers}. Such high intensities would also give rise to a novel regime of laser-matter interactions in which radiation reaction significantly impacts the particle dynamics. In this regime, the combination of the ultra-intense fields and the ultra-relativistic electrons generated by the laser would lead to copious emission of multi-MeV photons. The prospects of reaching this regime have stimulated numerous analytical and numerical studies on gamma-ray production at these high laser intensities~\cite{Brady12, Brady13, Brady14, Ridgers12, Nakamura, Duclous, Ji}. Nevertheless, one would have to wait until the intensities exceeding $10^{23}$ $\mathrm{W/cm^2}$ are achieved in order to implement any of the proposed photon generation schemes.

There are, however, several facilities with the capability of reaching intensities up to $5 \times 10^{22}$ $\mathrm{W/cm^2}$ within the  immediate future~\cite{TPW}. Most of the previous numerical studies have concluded that radiation reaction effects show little promise for converting an appreciable fraction of laser energy into high-energy photons at laser intensities below $10^{23}$ $\mathrm{W/cm^2}$ for PW class laser systems~\cite{Brady12,Nakamura}. It was concluded that an order of magnitude increase of either laser intensity or laser power would be required to achieve tens of percent for the total conversion rate.

\begin{figure*}[htbp]
	\centering
	\includegraphics[width=1\linewidth]{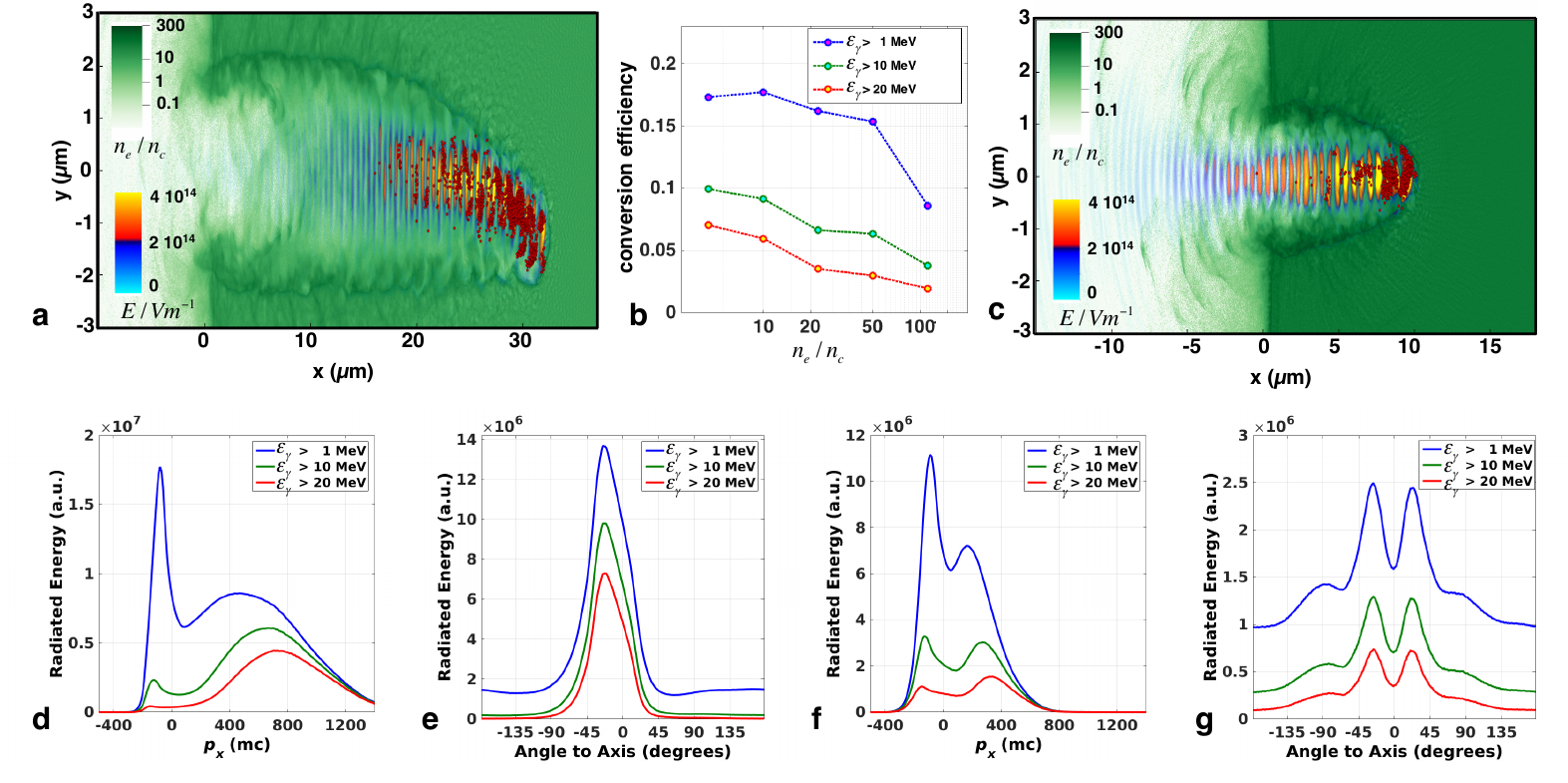}
	\caption{(color online) Results of 2D simulation density scan measuring the conversion efficiency of laser energy into gamma-rays (panel \textbf{b}) with energies above 1, 10, and 20 MeV. Panels \textbf{a} and \textbf{c} show density and electric field snapshots for $n_e=10n_c$ at $300$ fs  and $n_e=50n_c$ at $250$ fs, with photon emissions greater than 10 MeV in energy denoted by red circles.  The time-integrated radiated energies for these two simulations are shown as functions of emitting electron longitudinal momentum (\textbf{d} and \textbf{f}) and the angle of emission relative to the laser propagation axis (\textbf{e} and \textbf{g}).}
	\label{fig:conversion}
\end{figure*}

In this paper, we use 2D and 3D particle-in-cell (PIC) simulations to perform a numerical study of photon emission from laser-irradiated solid-density bulk targets using a PW-class pulse of intensity $5\times10^{22}$ $\mathrm{W/cm^2}$. We demonstrate that the quasi-static magnetic field generated by collective effects in a relativistically transparent laser-heated plasma greatly enhances the radiation reaction process, yielding tens of TW of directed MeV photons for a PW class laser system. Remarkably, the resulting laser energy conversion rate for multi-MeV photons is comparable to that previously predicted for an order of magnitude higher in intensity or power. Additionally, we propose a novel target geometry to control the directionality of this high-yield photon beam.


The synchrotron emission that is associated with the radiation reaction is expected to be the primary source of multi-MeV photons at the laser intensity of interest. The emission process of hard photons must be calculated self-consistently by explicitly accounting for the radiation reaction in the description of the electron dynamics. A probabilistic approach based on classical and QED synchrotron cross-sections, coupled with the subsequent reduction of the electron momentum~\cite{Duclous}, has been successfully implemented in the fully relativistic PIC code EPOCH~\cite{EPOCH,Ridgers14} that we use in this study. 

As the first step in our study, we examined how solid, thick targets perform when irradiated at normal incidence by a pulse with characteristics similar to those of the Texas Petawatt \cite{TPW}. We have performed 2D PIC simulations for several different target densities while using the same $5 \times 10^{22}$ W cm$^{-2}$ PW-class laser pulse. Specifically, we used a 1 $\mu$m wavelength, linearly polarized Gaussian pulse that is 100 fs in duration and that focuses to a spot 1.1 $\mu$m in radius. The peak normalized wave amplitude that corresponds to $5 \times 10^{22}$ W cm$^{-2}$ for this pulse is $a_0 \approx 190$. We initialized the targets as fully ionized uniform carbon plasmas with electron densities ranging from $n_e\approx4.5n_c$ and $n_e\approx110n_c$, which in practice corresponds to foam and plastic targets. Here $n_{c} = 1.1 \times 10^{21}$ $\mathrm{cm^{-3}}$ is the critical density. The cell size was 10 $\times$ 10 nm with  20 to 50 electrons and 10 to 20 ions per cell. 

The density decrease in the considered range describes the transition from the relativistically near-critical regime to the relativistically transparent regime. As shown in Fig. 1, for the higher densities the laser penetration into the target is only due to stable hole-boring \cite{Wilks, Robinson}, whereas at the lower densities the laser pulse propagates through the target due to the relativistically induced transparency \cite{Palaniyappan_2012}. The density scan shows that the  yield of multi-MeV photons increases with the onset of relativistic transparency. 

The relativistically-transparent targets demonstrate enhanced performance beyond just the overall yield of the multi-MeV photons. For the near-critical target, the photons are emitted by both forward and backward-moving electrons, as shown in Fig. 1f. This leads to a low degree of collimation (see Fig 1g). In the relativistically transparent regime, the emission pattern dramatically changes for higher energy photons. For energies above 10 MeV, the photons are emitted exclusively by forward moving electrons (panel \textbf{d}); this directly translates into a highly-collimated multi-MeV photon beam (panel \textbf{e}). The key features of relativistically transparent targets, then, are the increased efficiency and the dramatically enhanced collimation of the photon beam.

 However, these advantages are gained at the cost of the directivity of the photon beam. The laser beam propagation in the relativistically transparent regime is unstable due to the hosing instability \cite{Friou,Naseri}. As the laser pulse veers off its axis (see Fig 1a), so will the energetic electrons accelerated by the pulse. Since these electrons essentially emit parallel to their momentum, the directivity of the photon beam becomes just as unpredictable, as indicated in Fig.  1\textbf{e}. 


\begin{figure}[tb!]
	\centering
	\includegraphics[width=0.9\columnwidth]{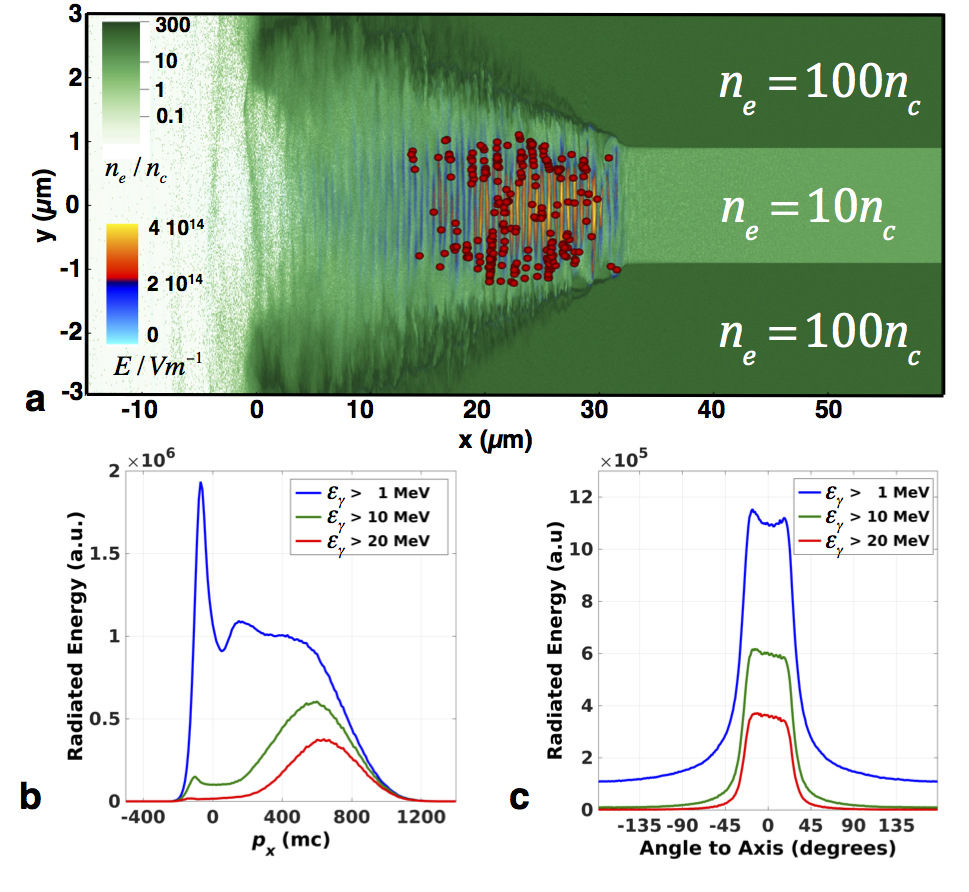}
	\caption{Density and electric field snapshot from the 2D channel simulation at $300$fs (\textbf{a}), where photon emissions greater than 10 MeV in energy are denoted by red circles. The time-integrated radiated energy in the simulation is shown as a function of emitting electron longitudinal momentum (\textbf{b}) and the angle of emission relative to the laser propagation axis (\textbf{c}).}
	\label{fig:channel}
\end{figure}

In order to utilize the best features of both regimes, we propose a novel target design shown in Fig. 2 that combines a relativistically near-critical (100 $n_{c}$) bulk target and a relativistically transparent (10 $n_{c}$) channel. Fig. 2b confirms that, similarly to the case of bulk relativistically transparent targets, the high energy photons are predominantly emitted by forward moving electrons. This leads to a well-collimated photon beam (panel \textbf{c}) with a high conversion efficiency of $\sim15\%$. What is more significant is that the near-critical bulk target guides the laser pulse so that the resulting photon beam is symmetric and directed along the axis of the channel. 

Still, the obtained result is counterintuitive, because it is well-known that energetic electrons co-propagating with the laser pulse in a vacuum are poor sources of gamma-ray emission. The acceleration induced by the laser is significantly higher for counter-propagating electrons~\cite{Blackburn,Vranic}. In order to understand the underlying mechanism of the gamma-ray emission in our setup, it is constructive to briefly examine the salient features of the synchrotron emission. The power emitted by a single electron is effectively determined by the acceleration in an instantaneous rest frame, $P_{rad} \propto \eta^2$, where 
\begin{equation}
\eta = \frac{\gamma}{E_S} \sqrt{ \left( {\bf{E}} + \frac{1}{c} {\bf{v}} \times {\bf{B}} \right)  - \frac{1}{c^2} \left( {\bf{E}} {\bf{v}} \right)^2}
\end{equation}
is a dimensionless parameter characterizing the acceleration. Here ${\bf{E}}$ and ${\bf{B}}$ are the electric and magnetic fields acting on the electrons, $\gamma$ and ${\bf{v}}$ are the relativistic factor and velocity of the electron, and $E_S \approx 1.3 \times 10^{18}$ $\mathrm{V/m}$ is the Schwinger limit. In our simulations, we typically observe $\eta^2 \approx 3\times10^{-3}$ during emissions of multi-MeV photons. For comparison, we first take the maximum electric field in the laser pulse, $E_0 \approx 6 \times 10^{14}$ $\mathrm{V/m}$, and the typical relativistic factor of emitting electrons, $\gamma \approx 700$. Not surprisingly, such a strong field produces tremendous acceleration for a co-propagating electron, with $\eta^2 \approx 0.1$. However, the magnetic field of the laser pulse counteracts the force from the electric field. As a result, the acceleration in the laser pulse drops to  $\eta^2 \approx 10^{-13}$ for an electron moving in the same direction as the pulse. We must note, though, that the electric field of the pulse does cause transverse oscillations, with a typical angle of 20$^{\circ}$ for the emitting electrons as they cross the channel's axis. Even in this case, $\eta^2 \approx 4 \times 10^{-4}$ and so is still an order of magnitude lower than what is observed in the simulations. The laser-plasma interaction itself must provide an additional contribution to the electromagnetic field that enhances the radiation reaction.

\begin{figure}[tb!]
	\centering
	\includegraphics[width=0.9\columnwidth]{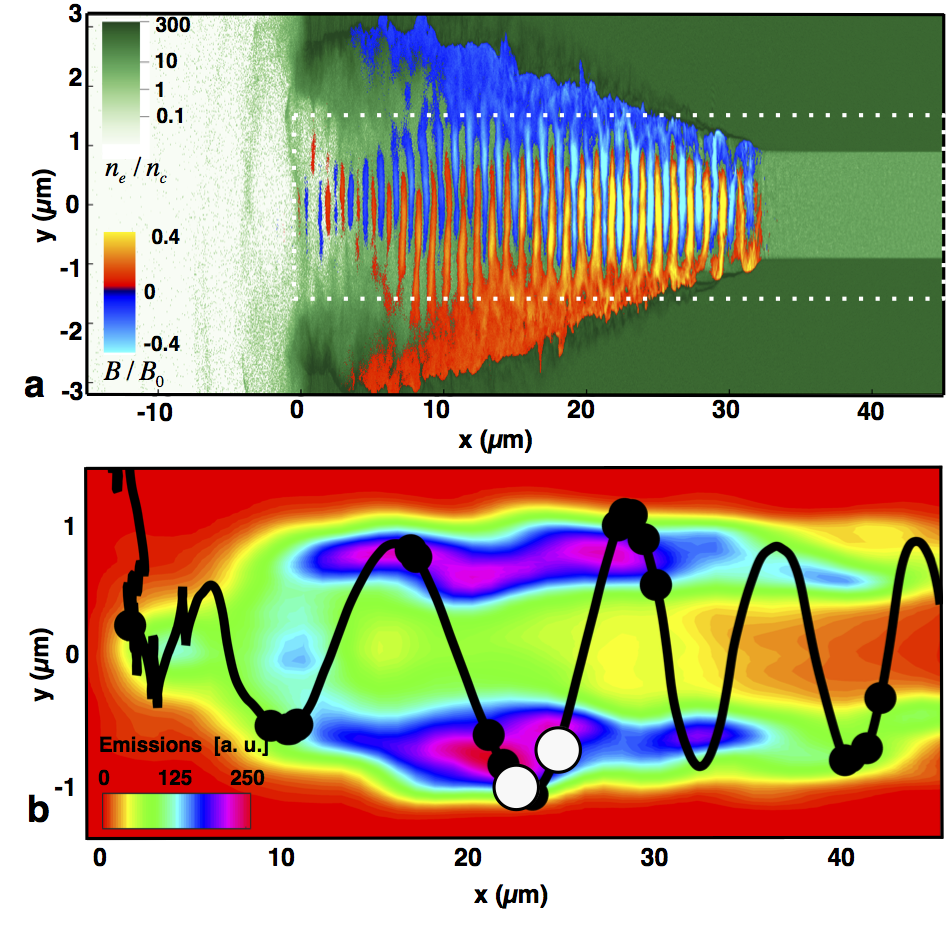}
	\caption{\textbf{a)} Magnetic field $B_z$ and electron density snapshot at $300$ fs from the 2D channel simulation. \textbf{b) }Sample emitting electron trajectory (black) showing the emissions (white) as it traverses the channel. The background color plot gives the emission count per cell for photons greater than 30 MeV.}
	\label{fig:bfield}
\end{figure}

A comparison of the electric (Fig 2a) and magnetic (Fig 3a) field snapshots in the channel reveals the presence of a strong magnetic field generated by the electron current, clearly visible near the edges of the channel. The maximum magnetic field that the channel electrons can generate can be estimated as
\begin{equation} \label{mag}
B \approx \frac{4 \pi}{c} |e| n_e c R \approx 6 \times 10^5 \mbox{ T},
\end{equation}
where we assume that the electron density $n_e$ is roughly the channel electron density and that $R$ is the channel radius. This expression further assumes that all of the electrons are moving forward with relativistic velocities, whereas in reality, some electrons move under an angle with respect to the channel axis. Though Eq.~(\ref{mag}) consequently overestimates the electron current and the corresponding magnetic field, it provides an order-of-magnitude estimate, $B \approx 0.3 B_0$, that is comparable to the $\sim 0.2 B_0$ that we observe in the simulation. Here $B_0 \approx 2 \times 10^6$ T is the magnetic field of the wave. We therefore can draw two important conclusions. First, the estimate indicates that the slowly-changing magnetic field is generated by the bulk electrons in the channel. Of greater significance, however, is that the plasma can  sustain a magnetic field whose strength is comparable to that of the laser pulse. 

The strong self-generated magnetic field considerably increases the acceleration experienced by the electrons moving along the channel. The crucial point is that the force exerted by such a magnetic field is not compensated by an electric field, as in the case of the laser pulse. For the characteristic strength of the self-generated magnetic field of $B \sim 0.2 B_0$, we have $\eta^2 \approx 4 \times 10^{-3}$, consistent with the values observed in the simulations. This attests to the enhancement of the gamma-ray emission in the channel being directly linked to the presence of a strong self-generated magnetic field.

We illustrate this effect of the self-generated field by analyzing  the emission pattern along a typical electron trajectory shown in Fig. 3\textbf{b} (black line). We chose an electron whose $\gamma$-factor during the emissions of multi-MeV photons is near the well-pronounced peak shown in Fig. 2\textbf{b}. After being injected into the channel, the electron is accelerated in the forward direction by the intense laser pulse. The laser electric field drives strong transverse oscillations, while the magnetic field generated by the plasma confines the electron inside the channel. The deflections by the magnetic field of the channel result in emissions of multi-MeV photons, and that is why most of the emissions occur in the vicinities of the turning points. This pattern is evident along the trajectory, where we show the locations of photon emissions with energies above 2 MeV and specifically highlight the photon emissions with energies above 30 MeV.   

Furthermore, the emission locations along this electron trajectory are representative of where the bulk of the radiating electrons emit hard photons, as evidenced by the locations of the photon emissions above 30 MeV given in panel \textbf{b}. The electrons emit at their turning points when they reach the strong quasi-static magnetic field at the edges of the channel, and thus the emissions are localized off axis instead of where the laser fields have the highest amplitude.


\begin{figure}[htb!]
	\centering
	\includegraphics[width=0.9\columnwidth]{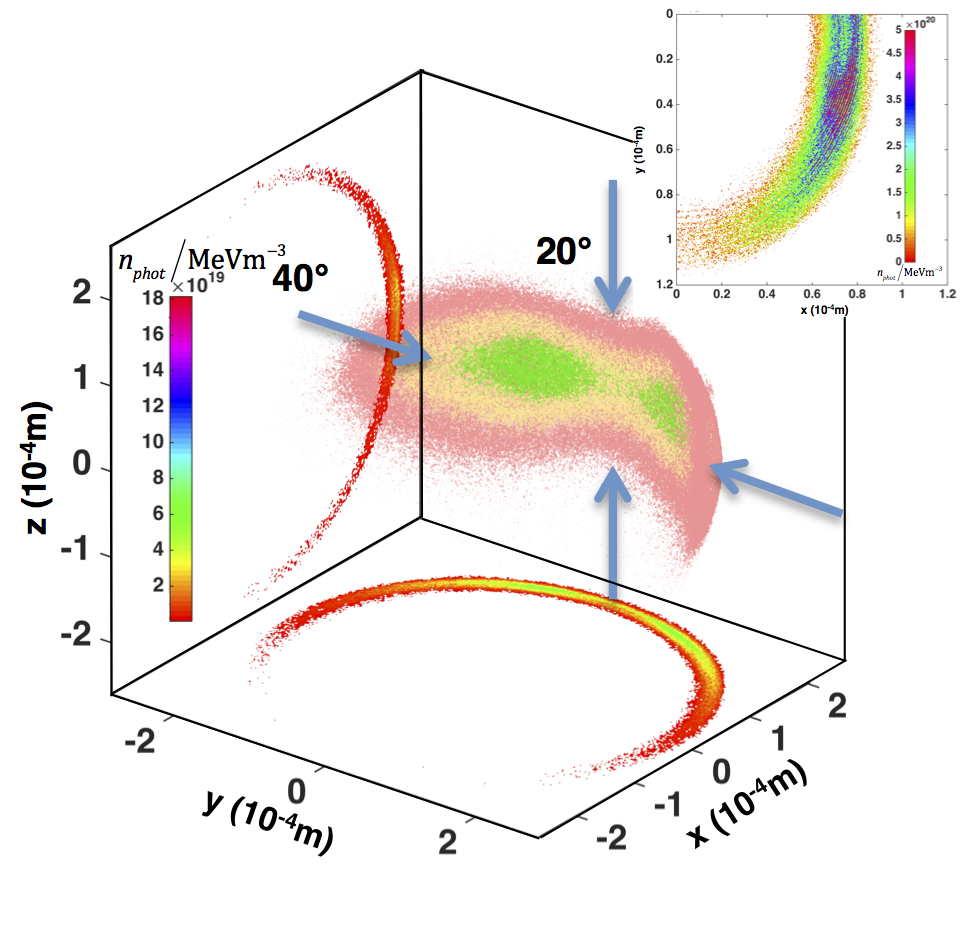}
	\caption{Volume plot of the photon energy density as iso-contours for 0.2 (green), 0.1 (light yellow) and 0.05 (light red) of the maximum density after 1 ps. On the left and bottom of the box are projections of the planes ($x$,0,$z$) and ($x$,$y$,0). The inset shows a central cross-section through the ($x$,$y$) plane, detailing the temporal substructure of the photon beam after 500 fs. }
	\label{fig:3d}
\end{figure}

 In order to make quantitative predictions for the energy conversion rate into multi-MeV photons, we performed a 3D simulation using the  same laser pulse parameters and the same target design, now with cylindrical symmetry. We used 23$\times$60$\times$60 nm cells with 50 electron and 25 ions per cell. The key qualitative features of the laser-target interaction are in good agreement with what has been observed in the 2D case. We recorded the locations and times of the emissions in the simulation and then propagated the photons, as a part of post-processing, for 1 ps to observe the far-field distribution (Fig 4). We find that the conversion efficiency into photons with energies above 1 MeV is reduced to 3.5\% in the 3D calculation, but the photon divergence out of the plane corresponding to the 2D simulation is significantly smaller than the divergence within that plane (i. e., in the $(x,y)$-plane). 

The photon beam is thus more powerful because of the anisotropic divergence than what one would expect when extrapolating the result of the 2D simulation. The beam is also much shorter than the laser pulse, with a characteristic duration of roughly 30 fs for photons with energies above 10 MeV. The photons are emitted in bursts, which leads to very pronounced beam intensity modulations a periodicity of the laser period. The total number of the multi-MeV photons in the beam is roughly $5.6 \times 10^{12}$. Using the calculated conversion efficiency and beam divergence, we expect that the cylindrical channel can effectively convert a PW-class 69 J laser pulse into a multi-TW beam of multi-MeV photons with the total energy exceeding 1 J.


In conclusion, we have examined photon emission from laser-irradiated bulk solid density targets at a laser intensity of $5 \times 10^{22}$ $\mathrm{W/cm^2}$ through the use of 2D and 3D PIC simulations. We have identified a regime in which a collimated multi-MeV photon beam with conversion efficiency comparable to those predicted well above $10^{23}$ $\mathrm{W/cm^2}$ can be generated even at the considered intensity. Relativistic transparency allows the laser pulse to propagate within the high density bulk of the target, generating both a beam of energetic electrons with GeV-level energies and consequently a strong Mega-Tesla level, slowly-evolving azimuthal magnetic field that significantly enhances the radiation reaction.  A preformed target geometry with a channel that becomes relativistically transparent during the interaction can control the photon beam directivity and yield tens of TW of directed MeV photons for a PW class laser. The properties of this emitted photon beam might enable the development of novel applications in areas of imaging, medical treatment, isotope production, and nuclear physics.

This research was supported by AFOSR Contract No. FA9550-14-1-0045. D.J.S. was supported by the DOE SCGF administered by ORISE-ORAU under Contract No. DE-AC05-06OR23100. Simulations were performed using EPOCH code (developed under UK EPSRC Grants No. EP/G054940/1, No. EP/G055165/1, and No. EP/G056803/1) using HPC resources provided by the TACC at the University of Texas. We also would like to thank Dr. Chris Ridgers for a helpful discussion of the radiation reaction module in EPOCH.


\end{document}